# In-Person, Hybrid or Remote? Employers' Perspectives on the Future of Work Post-Pandemic

**Divyakant Tahlyan**
Transportation Center, Northwestern University
Email: dtahlyan@u.northwestern.edu

**Hani Mahmassani**
William A. Patterson Distinguished Chair in Transportation,
Transportation Center
Northwestern University
600 Foster Street, Evanston, IL 60208, USA
Email: masmah@northwestern.edu
Tel: 847.491.2276

**Amanda Stathopoulos**
Department of Civil and Environmental Engineering, Northwestern University
A312 Technological Institute
Phone: +1 847-491-5629, Fax: +1 847-491-4011
2145 Sheridan Road, Evanston, IL, 60208, USA
Email: a-stathopoulos@northwestern.edu

**Maher Said**
Department of Civil and Environmental Engineering, Northwestern University
Email: MaherSaid@u.northwestern.edu

**Susan Shaheen**
Professor, Civil and Environmental Engineering
Co-Director, Transportation Sustainability Research Center
University of California, Berkeley
408 McLaughlin Hall
Berkeley, CA 94720
Email: sshaheen@berkeley.edu

**Joan Walker**
Department of Civil and Environmental Engineering
University of California at Berkeley
111 McLaughlin Hall,
Berkeley, CA 94720-1720, United States
Email: joanwalker@berkeley.edu

**Breton Johnson**
Transportation Center, Northwestern University
600 Foster Street, Evanston, IL 60208, USA
Email: bretj@northwestern.edu

## Abstract

We present an employer-side perspective on remote work through the pandemic using data from top executives of 129 employers in North America. Our analysis suggests that at least some of the pandemic-accelerated changes to the work location landscape will likely stick; with some form of hybrid work being the norm. However, the patterns will vary by department (HR/legal/sales/IT, etc.) and by sector of operations. Top three concerns among employers include: supervision and mentoring, reduction in innovation, and creativity; and the top three benefits include their ability to retain / recruit talent, positive impact on public image and their ability to compete. An Ordered Probit model of the expected April 2024 work location strategy revealed that those in transportation, warehousing, and manufacturing sectors, those with a fully in-person approach to work pre-COVID, and those with a negative outlook towards the impact of remote work are likely to be more in-person-centered, while those with fully remote work approach in April 2020 are likely to be less in-person-centered. Lastly, we present data on resumption of business travel, in-person client interactions and changes in office space reconfigurations that employers have made since the beginning of the pandemic.

**Keywords:** telework, employer survey, COVID-19, future of work, employer attitudes, choice models, latent class analysis

## 1. Introduction

The changing nature of (remote) work as a result of the COVID-19 pandemic in the United States (and globally) has been studied significantly over the past three years regarding how it evolved through the pandemic; how it may look in the near future and long term; what sort of impact it may have on cities; and how cities may prepare for these changing trends. There is significant emerging evidence from multiple datasets that a large proportion of the pandemic-accelerated telework trends are likely to stick well beyond the pandemic and that this will have major implications for urban mobility and the functioning of cities (Barrero et al., 2021b; Chapple et al., 2022; Conway et al., 2020; Javadinasr et al., 2021; Mohammadi et al., 2022; Ramani and Bloom, 2021; Salon et al., 2021; Tahlyan et al., 2022a; Tahlyan et al., 2022c).

Existing studies point to several factors that have and will continue to shape employee preferences regarding the extent to which they will continue to work remotely. On one end, studies point to several benefits for workers. These include travel time savings due to not needing to commute, improved productivity, well-being, and health outcomes, ability to provide care for other household members and dependents, and cost savings like transportation fuel cost of commuting or housing cost in cases where moving to a less expensive area is an option are some factors that employees might consider in evaluating the extent to which to continue to telework. On the other end, telework presents several challenges, such as a fuzziness of boundaries between work and life, lack of appropriate workspace and work environment at home like distraction from others, appropriate furniture and technology at home, and limitations in the ability to coordinate between teammates, opportunity for professional networking or career advancement, to mentoring and supervision opportunities (Aksoy et al., 2023; Barrero et al., 2021b; Dua et al., 2022; Hopkins and McKay, 2019; Kłopotek, 2017; Mannering and Mokhtarian, 1995; Owl Labs, 2022; Parker et al., 2020; Parker et al., 2022; Ramani and Bloom, 2021; Tahlyan et al., 2022c; Teevan et al., 2022).

A major issue with most existing research is that evolving remote work trends are only studied from the perspective of the *employees* even though any future trends are unavoidably a function of both *employee preferences and the employer-side decision* to allow remote work. Emerging literature points to arguments on both ends as to whether employers may continue to provide telework as an option. Evidence exists that productivity while working from home during the pandemic was equal to or better than expected for most (Barrero et al., 2021b; Owl Labs, 2022; Teevan et al., 2022) and this may be a factor that employers may consider to select remote work strategies. However, there is also evidence of differences between employee-perceived productivity in remote work settings compared to what employers perceive (Bloom et al., 2023).

There is also an argument of cost savings for the employers in terms of reduced office space requirement and not needing a cost of living adjustment to salaries, especially in cases where an employee works fully remotely from a less expensive city (GWA, 2021; Pratt, 1984). There is also growing evidence that employees are increasingly demanding more flexibility in work location and are even ready to either accept 5-10% pay cuts in order to have more flexibility or are ready to quit and switch to companies that offer better work location flexibility (Barrero et al., 2021a; Dua et al., 2022; Owl Labs, 2022; Parker et al., 2020). On the con side, there is evidence that remote work hurts collaborations and innovation; and this may be important for companies that rely on innovation to maintain a competitive edge over others (Yang et al., 2022). Similar to the case for productivity, also here there is a gap between the employee and employer expectations for remote work and this difference in expectations at the two ends may decide the eventual remote work adoption (Barrero et al., 2021b). Overall, while there are benefits of remote work for employees and potentially for employers too, there is still disagreement between the two parties on several matters and the result of this power struggle will potentially determine the future remote work landscape.

Even though employers are the ultimate decision-makers in this evolving remote work landscape, only a handful of studies have looked at the employer's perspective to understand the future of work (Alexander et al., 2021; Baker, 2020; Ozimek, 2020). However, these studies either focus on just the knowledge / information workers and do not present differences across various sectors or have surveyed human resources or mid-level managers only. Since the COVID-19 pandemic potentially proliferated telework beyond knowledge/information sectors and since ultimate company-wide telework policies are going to be driven by top-level executive decisions, we argue that a survey of a diverse set of top-level executives is necessary to gain a thorough understanding of the employers' perspective on telework in the post-pandemic world.

In this paper, we present results from data collected over a 5-wave longitudinal survey conducted amongst top executives (~90% vice president and chief executive officer level, ~10% director level) of 129 unique employers in North America to understand the evolution and future landscape of remote work through and beyond the pandemic. While our data is neither representative of the population of employers nor it is a large sample, our data is diverse in terms of organizational characteristics captured like annual revenue, number of employees, and sectors of operations, and is rich in terms of the information available and hence provides several interesting insights regarding the potential trajectory of the future of work. We specifically focus on the following four aspects.

*First*, for those employees for whom telework is possible, we analyze how the employers' approach to remote work varied over time since the beginning of the pandemic and what approach they expect to take in the future. Specifically, we collected data on the employers' approach to remote work at the following time points:

- September 2019 (pre-COVID)
- April 2020 (early phase of the pandemic)
- October 2021
- November 2021
- January 2022
- April 2022
- August 2022
- October 2022 (expected approach)
- April 2024 (expected approach)

We analyze this data to understand how the average work location across organizations varied (or is expected to vary) over time at an aggregate level and also identify differences across various sectors of operations, departments within the same organization, and how it varied for organizations with different remote work approaches pre-COVID (in September 2019) and in the early phase of the pandemic (in April 2020).

*Second*, we present and analyze data on the employers' opinion regarding the expected impact (positive, negative, or neutral) of remote work on various business aspects in a hypothetical scenario where a 2-days per week remote work policy is adopted by their organization. Business aspects include: the ability to recruit/retain employees, profitability, long-term viability, ability to compete, ability to innovate, public image, employee productivity, creativity, and supervision/mentoring. First, we descriptively present the employers' top concerns and perceived benefits related to the impact of remote work on business aspects, and this is followed by the estimation of a latent class model which divides employers into two latent classes representing their outlook (positive or negative) towards remote work. The latent class analysis is extended with the estimation of a latent class membership model to understand the association between the identified latent classes and the demographic characteristics of the organizations. This analysis helps us understand the important factors that may shape employer policies regarding remote work in the future.

*Third*, we dive deeper into understanding the remote work landscape in April 2024, four years after the beginning of the pandemic by estimating an ordered probit model that associates an organization's sector of operations, pre-COVID and early pandemic approach to remote work, and the employer's outlook towards the impact of remote work on business aspects with the expected April 2024 remote work approach. The model provides insights regarding which organizations are more likely to showcase higher/lower in-person presence in the future.

Fourth, we present data on the extent of resumption of business travel of over 50 miles and in-person client interactions at the time of various waves of the survey; and how employers have (or plan to) reconfigured their office spaces in light of the expected changes in employee work locations. Collectively, this data provides insights into the extent of the permanency of the altered work location landscape because of the pandemic.

The structure of the paper is as follows. Section 2 presents a review of the recent literature on emerging telework trends in the U.S. and across the world, the literature on the employee side and employer side perspectives on remote work through the pandemic. Section 3 describes the details of the 5-wave longitudinal survey used in this study followed by some descriptive statistics. Section 4 presents the analysis methodology, which is followed by the presentation of results in Section 5. The paper is concluded with a summary and key takeaway from this study, along with limitations and directions for future work.

## 2. Literature Review

### *2.1. Emerging telework trends*

Several available datasets and studies point to the changing telework landscape in the United States as a result of the COVID-19 pandemic. We briefly discuss these emerging trends from a few of these U.S. based datasets / studies regarding the extent to which teleworking rates have evolved through the pandemic, and then discuss how these trends differ across sectors, cities, and internationally.

At an aggregate level, several studies point to a significant shift toward remote work since the beginning of the pandemic. The Survey of Working Arrangements and Attitude (SWAA) by Barrero et al. (2021b) has been tracking the remote work frequency in the United States using a repeated cross-sectional dataset since May 2020 with over a hundred thousand respondents across all waves. Their data from Fall 2022 shows that about 30% of paid full days are worked from home each week and this number was about 60% in May 2020 and about 5% pre-COVID (based on American Time Use Survey). Their results also align with two other data sets, one collected by the U.S. Census Bureau named the Household Pulse Survey (United States Census Bureau, 2021) which started including a question on telework in June 2022, and another one is Google's Mobility Reports dataset (Google Inc., 2022). Two surveys done by Pew Research in October 2020 (Parker et al., 2020) and January 2022 (Parker et al., 2022), respectively, reported that about 20% of employed adults with remote work friendly jobs worked from home all or most of the time pre-COVID but this number was found to be 71% during October 2020 and 59% during January 2022. Another survey done in March/April 2022 by McKinsey and Company (Dua et al., 2022) reported that 58% of job holders in the United States can work remotely at least part-time. Lastly, data collected by Tahlyan et al. (2022b) reported the number of employed individuals working exclusively from the office to be 52% in 2019, but this number decreased to 17.5% in April 2020 and rebounded to only 31.4% in March 2022, about two years since the beginning of the pandemic.

At least two studies also provide an international perspective on telework trends (Aksoy et al., 2022; Lund et al., 2020). Global Survey of Working Arrangements (G-SWA) by Aksoy et al. (2022) consists of data from 27 countries collected in mid-2021 and early 2022. Their study reports the number of full-time paid days (conditional mean values after controlling for other factors, though their sample over-represents highly educated respondents in most countries) spent working from home, in the week when the data was collected, to vary significantly across countries with the highest number of 2.6 days in India, followed by 2.4 in Singapore, 2.2 in Canada, 2.1 in Malaysia, 2.0 in the United Kingdom and Australia. Countries with the lowest rates of remote work include South Korea (0.5 days), Egypt (0.7 days), and Taiwan (0.8 days). Lund et al. (2020) estimate the potential share of time spent working remotely for various countries and found it to be higher for advanced economies like UK, Germany, US, Japan, France, and Spain, compared to emerging economies like Mexico, China, and India since a large proportion of the workforce in emerging economies is skewed towards agricultural and manufacturing like sectors with require physical presence.

On the end of telework across various sectors, several studies (Barrero et al., 2021b; Lund et al., 2020; Tahlyan et al., 2022a; Tahlyan et al., 2023) point to it being higher in sectors like Finance and Insurance; Management, Professional, Scientific, and Technical Services; and IT and Telecommunications; and lower in sectors like Agriculture; Accommodation and Food Services; Construction; and Transportation and Warehousing. Using data collected in January 2023, Barrero et al. (2021b) reported that the number of days working from home in the last week was 2.29 for the Information sector, 2.15 from the Finance and Insurance sector, 1.96 for the Professional and Business Services, 0.69 for Retail Trade, 0.63 for Transportation and Warehousing, and 0.58 in Hospitality and Food Services. Tahlyan et al. (2023) report similar findings where they found that the individuals in Transportation and Warehousing were more likely to be fully in-person throughout the pandemic; individuals in Health Care sector were less likely to be fully remote; individuals in Information and Professional, Scientific and Technical Services were more likely to be hybrid or fully remote; and individuals in Finance and Insurance sector were more likely to be fully remote. Regarding post-pandemic trends in remote work, they also report individuals in Transportation and Manufacturing, Health Care, and Education sectors to be more in-person and individuals in Information, Finance and Insurance, and Professional, Scientific, and Technical Services sectors to be more remote in April 2024, about 4 years since the beginning of the pandemic. Lastly, they reported higher uncertainty in the April 2024 work location for those in the Information sector.

Lastly, Barrero et al. (2021b) and Chapple et al. (2022) provide insights into the variation in telework trends across cities. Barrero et al. (2021b) report the percentage of paid full-time days worked from home to be higher in the top 10 cities by population compared to other cities and show these trends to be consistent throughout the pandemic. Chapple et al. (2022) look at the downtown recovery data from various cities across the United States/Canada and found it to be lowest in cities like San Francisco, Portland, Indianapolis, and Seattle and this was correlated with the percent of jobs in sectors like Professional, Scientific and Technical Services in those cities, which other studies have shown to be high adopters of remote work.

### *2.2. Employee perspective*

Recent literature and data sources point to several factors that are continuing to shape employee preferences regarding expected remote work adoption in the future including productivity while working from home, commute travel time savings due to remote work, ability to care for other household members, and children, etc. Here, we briefly discuss emerging evidence regarding the employee perspective on remote work in terms of their experiences through the pandemic and their expectation post-pandemic regarding telework. These factors will potentially shape how employers form their perspectives and decide on future telework policies.

Most sources suggest that the remote work experiences during the pandemic were positive for most individuals due to several contributing factors. On the end of overall satisfaction with telework, using data from 318 working adults in the United States, Tahlyan et al. (2022c) report that about 74.21% of individuals in their data were either or would have been (if telework was an available option) somewhat or very satisfied with telework during the pandemic. Several sources also report telework during the pandemic didn't impact the productivity of the employees, and in fact, many report increased productivity due to remote work. For example, as Barrero et al. (2021b) note, the work productivity experiences during the pandemic were largely positive for the employees. From the data collected from about 30,000 respondents in the United States, they report that only 13.9% of individuals reported worse than expected productivity levels, 26.7% reported about the same level of productivity than before and the rest 59.5% reported better, substantially better or hugely better level of productivity than before. Another report by Owl Labs and Global Workplace Analytics (Owl Labs, 2022) with data from over 2300 full-time U.S. workers reported that 62% of the workers feel more productive when working remotely, with 66% of the millennials feeling more productive while working from home while this number is only 46% for boomers. A 2021 version of this survey with data from 2050 full-time U.S. workers reported that about 90% of individuals felt equally or more productive working from home. However, Teevan et al. (2022) pointed out that measuring productivity may be difficult due to definitional differences and may also be impacted by most of these numbers being self-reported. They also pointed to a large difference between productivity gains reported by employees and the employee productivity gains reported by their employers.

Several other factors played a role in either overall satisfaction with telework or in terms of their impact on the productivity of working from home. Tahlyan et al. (2022c) reported that productivity gains, travel time savings due to not needing to commute, and quality of life improvements contributed to the experienced benefits of telework and were positively associated with satisfaction with telework during the pandemic. They also found that lack of appropriate technology and distractions from other members of the households were some of the experienced barriers to teleworking and contributed negatively to telework satisfaction. Lastly, they also report that satisfaction was lower for those who were younger or older in age groups or those living alone, potentially due to loss of networking opportunities, difficulty in mentoring and supervising employees, and loss of visibility for career advancement. Barrero et al. (2021b) and Aksoy et al. (2023) also talk about commute time savings as a factor that makes telework valuable to employees. Barrero et al. (2021b) found out that the commute length was longer for higher-earning employees so they might value telework more than others. Aksoy et al. (2023) quantified the commute time savings associated with working from home based on data from 27 countries to be 72 minutes in their sample, with values in the range of 100 minutes for individuals in China, India, and Japan; around 70-80 minutes for Australia, Austria, Brazil, Russia, United Kingdom and South Korea; and 55 minutes for United States. They also found out that about 40 percent of the time saved gets allocated to work and about 11 percent is allocated to caregiving activities. Along the lines of career advancement and visibility to managers, Owl Labs (2022) reports that about 49% of respondents feel that their managers view office-going employees to be more hardworking and trustworthy than others. They also report that hybrid workers save about $19.11 each day due to remote work, a majority of which comes from commute expenses and expenses on lunch. They also mention feeling disconnected from others during hybrid/remote meetings and internet problems to be some of the contributing factors impacting the telework experiences of employees. Similar factors were also identified by Teevan et al. (2022), Parker et al. (2020), and Parker et al. (2022).

Literature also suggests changed expectations by the employees regarding remote work which will play an important role as employers decide on their remote work policies in the future. Owl Lab's 2021 state of remote work report (Owl Labs, 2022) reported that 65% of employees in their data expect some form of remote work going forward, with 1 in 3 ready to quit their jobs if they could not work remotely after the pandemic. Their report's 2022 version stated that 66% of workers would start looking for another job that would offer flexibility if the ability to work remotely is taken away. They also reported that 52% of the employees would take a pay cut of 5% or more and 23% would take a pay cut of 10% or more to have more flexibility. Dua et al. (2022) reported flexible working arrangement is one of the top motivators for finding a new job for employees. Parker et al. (2020) report that only 11% of employees with telework-friendly jobs want to work rarely or never from home. Parker et al. (2022) in a follow-up study reported that while the COVID-19 contagion risk was a motivator for individuals to work from home early in the pandemic, most employees now cite personal preferences for doing so like childcare responsibilities. Barrero et al. (2021a) found out that 40% of the employees with the ability to work from home at least once a week are ready to seek a new job if they are recalled back to the office full-time.

### *2.3. Employer perspective*

A limited number of studies exist that present the employer's perspective of remote work. We present a brief review of the emerging insights from these studies regarding employers' experiences with remote work during the pandemic and their plans for the future.

A report from McKinsey and Company (Alexander et al., 2021) surveyed 100 C-suite, vice president and director-level executives from around the world and found that 9 out of 10 organizations are planning to take a hybrid approach to work going forward however most do not have a detailed plan on the implementation aspect. Most respondents also reported an improvement in productivity, customer satisfaction, employee engagement, diversity, and inclusion, however, the productivity improvements were higher for those who kept their employees connected. They also found out that organizations, where productivity improved, were those who trained their managers to deal with situations like providing feedback to employees in an online setting and those who are continuously experimenting with different strategies regarding remote work. The organization leading in productivity improvements also reimagined their hiring processes with more and more recruiting events being held remotely.

Another study by Good Hire, a background check service for businesses (Korolevich, 2022) surveyed 3500 managers in the United States to understand their remote versus in-person preferences for their employees and found out that 75% of the managers preferred at least some form of in-person work, and 60% of the managers agreed or strongly indicated a full-time in-person work in the near future. They also found that 73% of managers agree that productivity and engagement has improved or remained the same during the pandemic and that 68% of managers agree that fully remote operation would add or not impact their profits.

Gartner Inc. (Baker, 2020), a technology research and consulting firm, consisting of 127 company leaders found that 82% of them plan to allow some form of remote work amongst their employees, with 47% allowing full time remote work.

A study done in the early phase of the pandemic is by Ozimek (2020) who collected data from 1500 hiring managers including executives, VPs, and managers. They found out that a majority of managers agree that remote work functioned somewhat to much better than expected during the pandemic with no commute, reduced non-essential meetings, fewer distractions, and increased productivity being some of the main benefits. About one-third of managers also reported reduced team cohesion, difficulty in communications, and disorganized teams as some of the aspects that worked poorly during the pandemic. 61.9% of the managers reported more remote work than before going forward with 21.8% reporting an entirely remote workforce.

## 3. Survey and Sample Description

### *3.1. Survey Description*

The data used in this study comes from a 5-wave longitudinal survey conducted between October 2021 and August 2022 amongst top executives of 129 unique North American companies. Initial invitations consisted of a total of 198 unique employers, primarily associated with Northwestern University Transportation Center (NUTC)'s business advisory council (BAC) (NUTC, 2022). Upon invitation, respondents had an option to either opt out of the current and future surveys, complete the current survey but opt out of future surveys or to complete the current survey and express interest in being invited for the future surveys. Based on the respondent's responses to this question, they were re-invited for future surveys.

The survey was designed via the Qualtrics platform and was sent to the respondents via email. The timelines for various waves of the survey are given below:

- Wave 1: October 2021
- Wave 2: December 2021
- Wave 3: January 2022
- Wave 4: March 2022
- Wave 5: August 2022

Five categories of questions were asked of the respondents regarding their organization:

**1)** *demographic information*, which included questions on:
   a. the sector of operations of the organization defined as per the North American Industry Classification System (NAICS) (Office of Management and Budget, 2022) with an option to also describe their specific sector in their own words
   b. percentage of the workforce that is typically customer-facing and/or must perform their work on-site (to be expressed using a sliding scale varying from 0 to 100)
   c. number of employees in their organization (options included: 1-9, 10-49, 50-99, 100-499, 500-999, 1000-4999, 5000-9999, 10000 or more)
   d. annual revenue of their organization (options included: less than $10 million, $10 million - $100 million, $100 million to $500 million, $500 million - $1 billion, more than $1 billion, prefer not to disclose)
   e. the region of operations (option included (select all that apply): Northwest, West, Midwest, Southwest, Southeast, Mid-Atlantic, Northeast, Nationally across the U.S., Internationally).

   These questions were *asked* in every wave of the survey.

**2)** *employers' approach to employee remote work* for whom it is possible at various time points before and during the pandemic and the expected work location in the future. The time points included:
   a. *September 2019* (before the pandemic)

b. *a few past time points during the pandemic relative to when the survey was done* (Wave 1: none, Wave 2: April 2020[1], Wave 3: April 2020 and October 2021[2], Wave 4: April 2020, October 2021, and January 2022, Wave 5: April 2020, October 2021, January 2022 and April 2022)
c. *current time point* (Wave 1: October 2021, Wave 2: November 2021, Wave 3: January 2022, Wave 4: April 2022, and Wave 5: August 2022)
d. *few future time points relative to when the survey was conducted* (January 2022[3], April 2022, October 2022 and April 2024)

The possible response options included: Fully in-person, Mostly in-person, About 50/50, Mostly remote, Fully remote and I don't know[4]. For the future time points in each wave, an additional option of 'wait and see' was also included to capture the responses where an employer has not yet decided on their future remote work policy. Given that the remote work approach potentially varies by department, we inquired separately for the following three departments within an organization (if applicable): sales / marketing, IT / development, administration / finance / legal / human resources.

**3)** their opinion towards the potential impact of a 2-day-a-week remote work policy in their organization on various business aspects. Specifically, we asked the employers to imagine that their company committed to a future work program allowing a hybrid workforce with an option of remote work for 2 days a week and what effects will such program have on the following 9 business aspects:

a. ability to recruit / retain employees
b. profitability
c. long-term viability
d. ability to compete
e. ability to innovate

---

[1] This time point was included in wave 2 and beyond to capture the remote work policies during the early peak of the pandemic.
[2] Note here that October 2021 was a past time point in wave 2 and beyond but was current time point in wave 1. Similarly, January 2022 was past time point in wave 4 and 5 but was current time point in wave 3; and April 2020 past time point in wave 5 but was current time point in wave 4.
[3] Note here that January 2022 and April 2022 were future time points during waves 1 and 2. However, during wave 3 while January 2022 was current time point, April 2022 was a future time point. Similarly, during wave 4, April 2022 was a current time point.
[4] While we provided the respondent to choose 'I don't know' as an option in accommodate cases where the respondent does not know the approach their organization took or will take (potentially due to the respondent not being in a decision-making position), no respondent chose this option in any of the waves.

f. public image
g. employee productivity
h. employee creativity
i. employee supervision and mentoring

The possible response options included: very negative, somewhat negative, neither negative nor positive, somewhat positive, very positive, and no opinion. This question was asked in every wave of the survey.

**4)** additional questions on:

a. the extent of resumption of business travel of over 50 miles and in-person client interactions at the time of various waves of the survey; Specifically, we asked the employers to report the extent to which (in terms of percentage) business travel of over 50 miles and in-person client interactions in their organization has resumed compared to the pre-pandemic levels, with zero percent being that none has resumed and a hundred percent being that all have returned. This question was included in all waves of the survey.

b. whether their organization has (or plan to) added, reduced, relocated office space in the same or different area or building since the beginning of the pandemic. This question was only included in wave 5.

Table 1: Sample statistics for data in various waves and the combined data

| Variable | | All waves | Wave 1 | Wave 2 | Wave 3 | Wave 4 | Wave 5 |
|---|---|---|---|---|---|---|---|
| Number of Respondents | | 129 | 62 | 43 | 34 | 45 | 56 |
| Number of Employees | 1 to 99 | 14.00% | 10.30% | 9.10% | 8.80% | 11.60% | 13.70% |
| | 100 to 999 | 28.70% | 29.40% | 27.30% | 38.20% | 34.90% | 31.00% |
| | 1000 to 9999 | 22.50% | 29.40% | 29.50% | 23.50% | 20.90% | 15.50% |
| | 10000 or more | 34.90% | 30.90% | 34.10% | 29.40% | 32.60% | 39.70% |
| Sector | Transportation | 53.50% | 51.50% | 54.50% | 64.70% | 44.20% | 58.60% |
| | Manufacturing | 7.80% | 10.30% | 11.40% | 8.80% | 9.30% | 8.60% |
| | Others | 38.80% | 38.20% | 34.10% | 26.50% | 46.50% | 32.80% |
| Annual Revenue | Less than $100 mil | 24.80% | 17.70% | 11.40% | 20.60% | 18.60% | 31.00% |
| | $100 mil to $1 bil | 16.30% | 22.10% | 22.70% | 29.40% | 25.60% | 13.80% |
| | $1 bil to $5 bil | 27.10% | 32.40% | 31.80% | 23.50% | 25.60% | 20.70% |
| | More than $5 bil | 27.90% | 26.50% | 29.50% | 26.50% | 25.60% | 31.10% |
| | Prefer not to disclose | 3.90% | 1.50% | 4.50% | 0.00% | 4.70% | 3.40% |
| Percentage of workforce required to in-person | Percentage | 55.48% | 54.03% | 59.45% | 55.94% | 49.93% | 56.59% |

Our survey also included questions on policies regarding masking, vaccinations, testing for COVID-19 infection at the work location and the work location policies their peer companies have adopted, return of business travel of over 50 miles, local in-person client interactions, and firm's office space reorganizations decisions. However, for brevity, we have not included results from the data corresponding to these sets of questions. Most of these questions were asked in every wave of the survey but a few were included in one or two specific waves taking into consideration the evolving telework trends at the time of the survey.

*3.2. Sample Descriptives*

Table 1 presents the descriptive statistics for the data corresponding to each wave of the survey as well as for the combined data. Figure 1 present the distribution of region of operations of various organizations in the data.

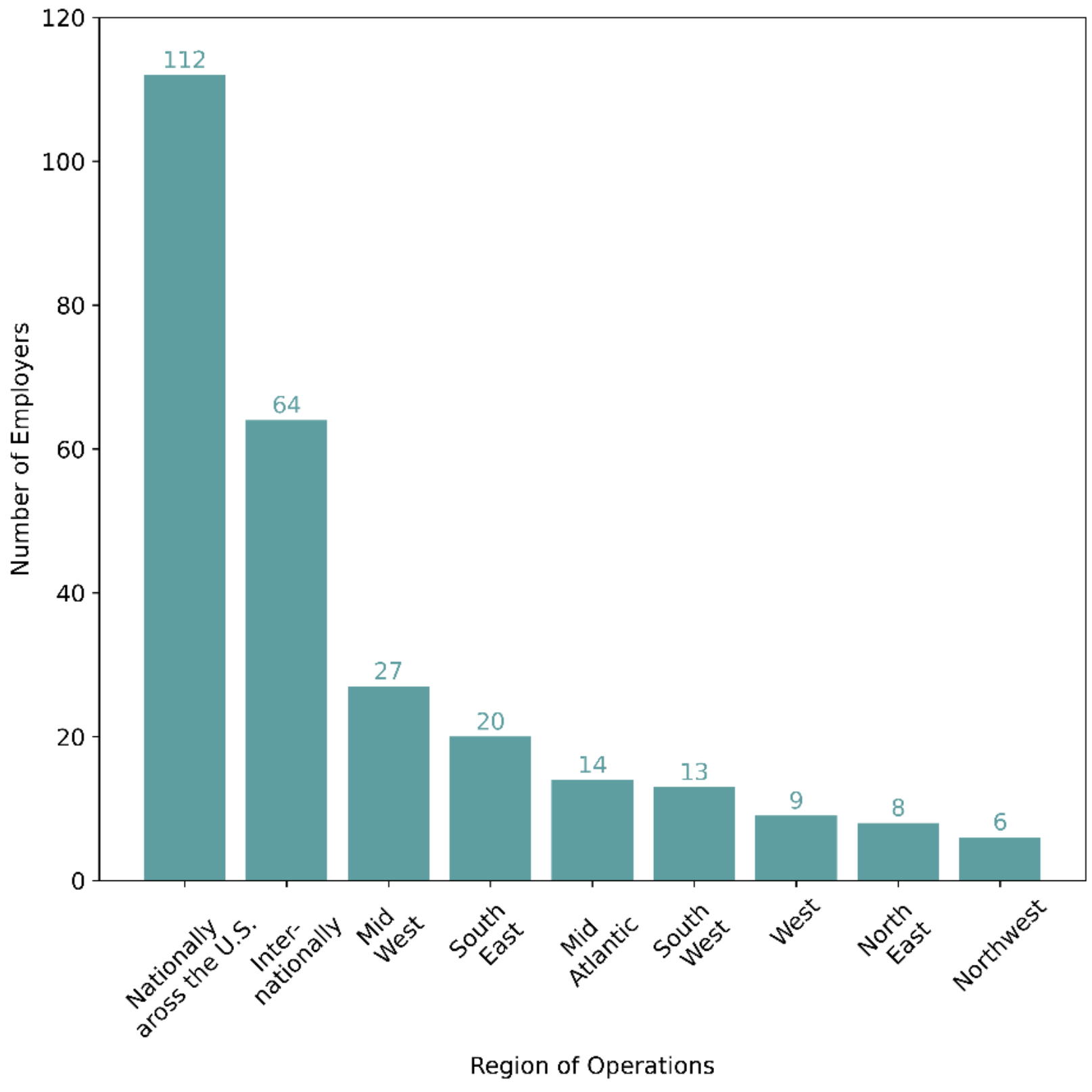

Figure 1: Employers' organizations' region of operations

In terms of the sample size, our data correspondent to 129 unique employers across the 5 waves who completed the survey in at least one of the waves, where 62 employers completed the survey in wave 1, 43 completed it in wave 2, and 34 completed it in wave 3. In wave 4 and 5, we augmented our invitation list with new employers to boost the sample size and this led to 45 respondents completing the survey in wave 4 and 56 respondents completing the survey in wave 5. Overall, we send invitations to 198 unique employers. Across the 5 waves, 71 employers responded to only one wave, 29 responded to two waves and another 29 respondents to three or more waves of the survey.

Our sample consists of about 61% of employers from the transportation, logistics, warehousing and manufacturing sectors and the rest 39% from other sectors (which is because members of the NUTC's BAC are primarily transportation, logistics, and warehousing companies). Figure 2 presents a word cloud of the sector of operations of various respondents in the survey in their own words, which highlights that the transportation sector is prominent in the data. Other sectors included employers from freight, railroad, third-party logistics, software, consulting, management, and data services.

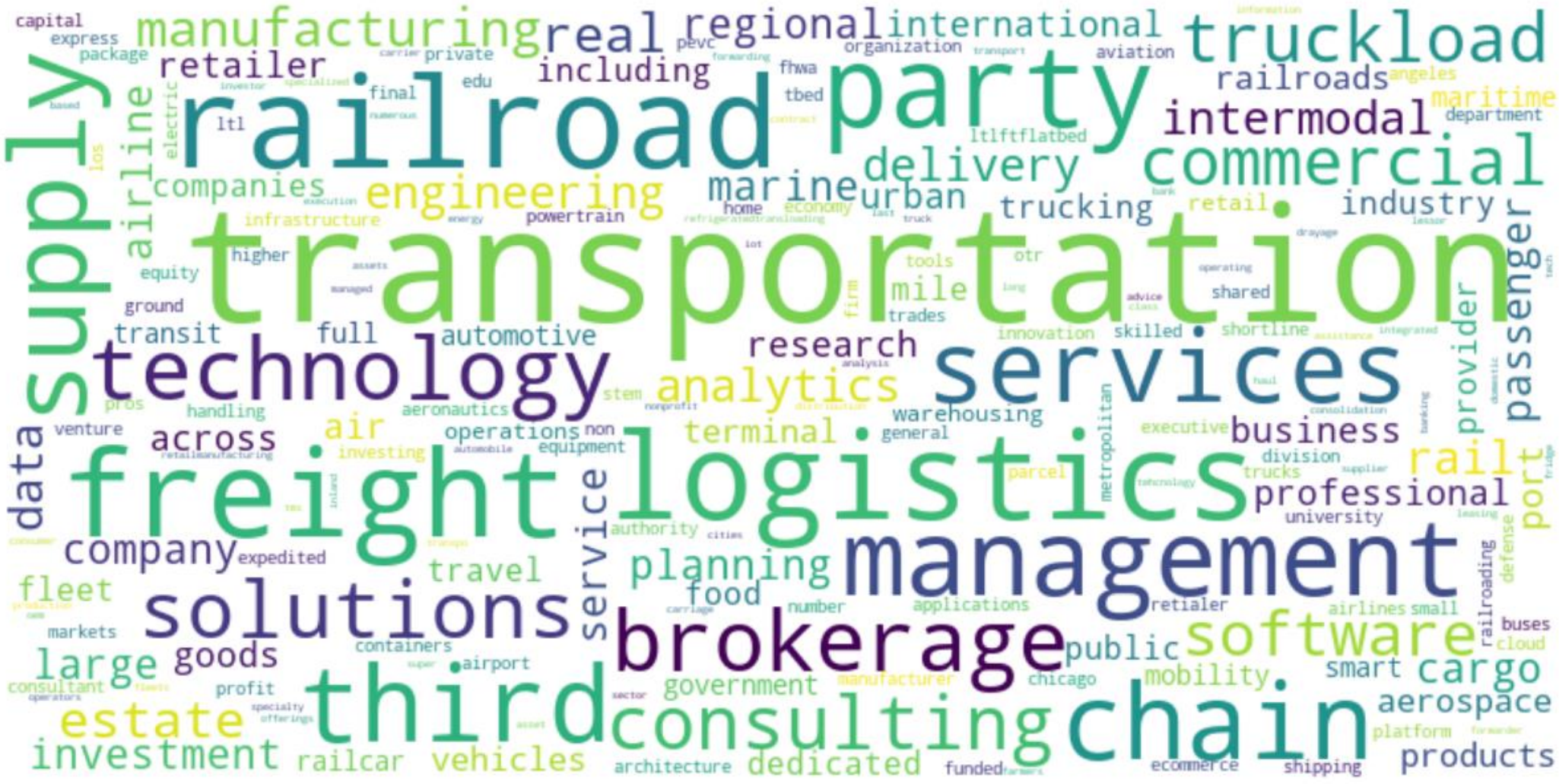

Figure 2: A word cloud of employer report (in their own words) description of the sector of operations of their company / organization

On an average, 55.48% of the workforce of the organizations in our data is customer-facing or must perform work in person with the standard deviation being 29.53%. A large share of companies in our sample have more than 10,000 employees, have more than $5 billion in annual revenue ($25 billion in many cases), and a majority have a presence across the United States (with several having an international presence). More than 90% of respondents in our data are vice president (VP) or chief executive officer (CEO) level senior executives, with the rest being director-level managers.

Overall, this highlights that the respondents in our data speak for a large number of employees and their decisions regarding remote work will potentially have an impact on the work location decisions of a large number of individuals.

## 4. Analysis Approach

### *4.1. Employers' approach to employee remote work for whom it is possible*

To analyze the data on the employer approach to employee remote work at different time points, we convert the 5-point Likert scale data ranging from fully remote to fully in-person to a numerical value ranging from 1 to 5 (1=fully remote and 5 = fully in-person)[5] and calculate average work location across the organization for different time points and derive the corresponding confidence intervals using the bootstrapping technique (Efron, 1992). Specifically, for a time point, we take the available work location approach data to estimate the average work location across organizations and then derive a 90% confidence interval around mean values using 500 resamples and the sample size per sample as the number of observations using the *sur* package in R (Harel et al., 2020). We present this data in several different ways:

- aggregate level (where averages are calculated across all organizations),
- across the sector of operations (transportation / manufacturing and others),
- across different departments (IT, sales, HR / admin / legal / finance),
- by different levels of remote work adoption in September 2019 (fully in-person, mostly in-person, and others)
- by different levels of remote work adoption in April 2020 (fully remote, mostly remote, and others).

[5] Cases where a respondent reported a 'wait and see' approach for the future time points were removed for this particular analysis since those response do not fit in the ordered scale of 1 to 5. However, in most cases they comprise of less than 10% of total number of observations available for a time point. Specifically, there were 1, 3, 10 and 15 responses with a 'wait and see' response for January 2022, April 20222, October 2022 and April 2024 time points, respectively. We elaborate more on these in the results section but these responses have been kept out for this analysis and of modeling due to limited sample size. Of course, the ultimate landscape of remote work will depend on the approach that these employers end up taking.

Given that some employers responded in multiple waves and hence their responses for certain time points are available more than once, we retain their oldest available response for time points which were in the past compared to when the survey was conducted, and retained their newest response in cases a time point was in the future compared to when the survey was conducted. This helps to minimize recall bias (Spencer et al., 2017) and corrects for updating the preferences of the employers. Along with the aggregate level data of work location across time, we also present wave-specific data for future time points, which captures changing employers' preferences regarding remote work, potentially impacted by employee pushback regarding return to office as well as changing pandemic situation since some of our data was collected at the height of the Omicron wave.

### *4.2. Employers' opinion on the impact of remote work on business aspects*

To analyze the data on the employer's view of the impact of 2-days a week remote work policy on various business aspects, we first analyze the data descriptively to highlight employers' top concerns and perceived benefits related to the impact of remote work on business aspects and later estimate a latent class model to stochastically divide employers into latent classes reflecting their outlook towards remote work. A base latent class model is first estimated without exogenous variables (like the sector of operations) and this is followed by the estimation of a class membership model where latent classes are associated with the sector of operations of the employer's organization[6]. The original data which was on a 5-point Likert scale ranging from Very negative to Very positive was converted to a 3-point scale to reduce complexity given a smaller sample size (positive, neutral, and negative) for both descriptive analysis as well as latent class analysis.

In a latent class model (see Collins and Lanza (2009) for more details), we used the 9 available business aspects (which are the ability to recruit / retain employees, profitability, long-term viability, ability to compete, ability to innovate, public image, employee productivity, employee creativity, and employee supervision and mentoring) as indicators to define the latent classes. For an unconditional latent class model (i.e. without exogenous variables), consider, $j = 1, \dots, J$ observed variables / indicators (9 business aspects in our case) with $r_j = 1, \dots, R_j$ response categories each (3 for each of the 9 business aspects, positive, neutral, and negative) form a vector of observed response pattern for each respondent, $\boldsymbol{y} = (r_1, \ \dots, \ r_J)$. We are interested in $P(Y = \boldsymbol{y})$, where $Y$ is a vector of response pattern and $\sum P(Y = y) = 1$. Two sets of parameters are estimated here: 1) $\gamma_c$, which represents the percentage of respondents in the data belonging to a latent class $c$; 2) $\rho_{j,r_j|c}$, which represents the probability of responding $r_j$ to the $j^{th}$ observed variable conditional on class membership. The response vector probability conditional on latent class $c$ in the model is written as:

$$P(Y = \boldsymbol{y}|L = c) = \prod_{j=1}^{J} \prod_{r_j=1}^{R_j} \rho_{j,r_j|c}^{I(y_j=r_j)} \tag{1}$$

where the fundamental expression to be estimated can be derived using the total probability theorem as:

$$P(Y = \boldsymbol{y}) = \sum_{c=1}^{C} \gamma_c \prod_{j=1}^{J} \prod_{r_j=1}^{R_j} \rho_{j,r_j|c}^{I(y_j=r_j)} \tag{2}$$

and the class membership probability for each respondent conditional on their response vector can be estimated using the Bayes' theorem as below:

$$P(L = c|Y = \boldsymbol{y}) = \frac{P(Y = y|L = c)P(L=c)}{P(Y=y)}$$

$$P(L = c|Y = \boldsymbol{y}) = \frac{\prod_{j=1}^{J} \prod_{r_j=1}^{R_j} \rho_{j,r_j|c}^{I(y_j=r_j)} \cdot \gamma_c}{\sum_{c=1}^{C} \gamma_c \prod_{j=1}^{J} \prod_{r_j=1}^{R_j} \rho_{j,r_j|c}^{I(y_j=r_j)}} \tag{3}$$

[6] Here, we also tried including other exogenous variable like annual revenue and number of employees but did not find them to be statistically significant.

The estimation of a membership model to incorporate exogenous variables $(x)$ in the model, $\gamma_c$ can be expressed as a function of $x$ and the fundamental expression conditional on $x$ can be written as:

$$P(Y = y \mid x) = \sum_{c=1}^{C} \gamma_c(x) \prod_{j=1}^{J} \prod_{r_j=1}^{R_j} \rho_{j,r_j|c}^{I(y_j=r_j)} \tag{4}$$

where instead of estimating $\gamma_c$, a logit formulation is used and $\beta$ parameters are estimated.

$$\gamma_c(x) = \frac{e^{\beta_0+\beta_{1c}x}}{1+\sum_{c'=1}^{C-1} e^{\beta_0+\beta_{1c}x}} \tag{5}$$

Model estimation was done using the *'poLCA'* package (Linzer and Lewis, 2011) in R programming language. We first estimate a model without covariates to determine the number of latent classes, where we varied the number of latent classes from 1 to 8 and then used the Bayesian information criterion (BIC) to pick the best model. To search for a global maximum instead of a local one, we tested 100 different starting values for each number of latent classes and picked the model with the greatest log-likelihood. After determining the number of latent classes, we re-estimated the model incorporating various exogenous variables, establishing that the sector of operations (transport / manufacturing or others) was the only significant variable.

*4.3. The future remote work landscape*

To understand the expected future landscape of remote work, we use the data related to the expected work location policy in April 2024 to estimate an Ordered Probit model. Considering the employer's April 2024 work location strategy on a 4-point ordered scale (1=Fully or mostly remote, 2=About 50/50, 3=Mostly in-person, 4=Fully in-person)[7], the model associates an organization's sector of operations, pre-COVID and early pandemic approach to remote work, and the employer's outlook regarding the impact of remote work on business aspects with the expected April 2024 remote work approach. An ordered probit model consists of a latent variable $y^*$ such that:

$$y^* = z\gamma + u \tag{6}$$

where $z$ is a vector of exogenous variables (covariates), $\gamma$ is a vector of estimable parameters and $u$ is the standard normally distributed error term. The latent propensity function $y^*$ is related to the reported $J-point$ response item $y$ (4-point in our case) in the following manner:

$$y = \begin{cases} 1 & \text{if} \quad y^* \leq \psi_1 \\ j & \text{if} \quad \psi_{j-1} < y^* \leq \psi_j \ \forall\, j \in (2, \dots, J-1) \\ J & \text{if} \quad \psi_{J-1} \leq y^* \end{cases} \tag{7}$$

where $\psi_j$ $(j = 1, 2, \dots, J-1)$ are estimable thresholds dividing the propensity equation. Given the above equations, probability $P(y)$ is:

[7] As mentioned earlier, 15 out of available 121 responses for this question were reported as 'wait and see' and those have been removed from the modeling analysis due to sample size. However, we present descriptive information on these responses in the results section. Further, while the original data available here is on a 5-point scale, we combine the responses from the fully remote and mostly remote responses categories since only one responded reported taking fully remote approach in April 2024.

$$P(y) = \begin{cases} \Phi(\psi_1 - z'\gamma) \\ \Phi(\psi_j - z'\gamma) - \Phi(\psi_{j-1} - z'\gamma) \ \forall\, j \in (2, \dots, J-1) \\ 1 - \Phi(\psi_{J-1} - z'\gamma) \end{cases} \tag{8}$$

where $\Phi(\cdot)$ is the standard normal cumulative distribution. We estimated this model using the *'lavaan'* package (Rosseel, 2012) in R programming language. For more information on ordered models, the readers are referred to Washington et al. (2020).

### *4.4. Business travel, in-person client interactions, and office space reorganization*

We descriptively analyze the data on the resumption of business travel of over 50 miles and in-person client interactions and office space reorganizations; present bar plots for each. Regarding business travel over 50 miles and in-person client interactions, we calculated the average percentage resumption across data from each wave of the data collection. This is followed by a presentation of data on office space reorganization strategies adopted by the employers.

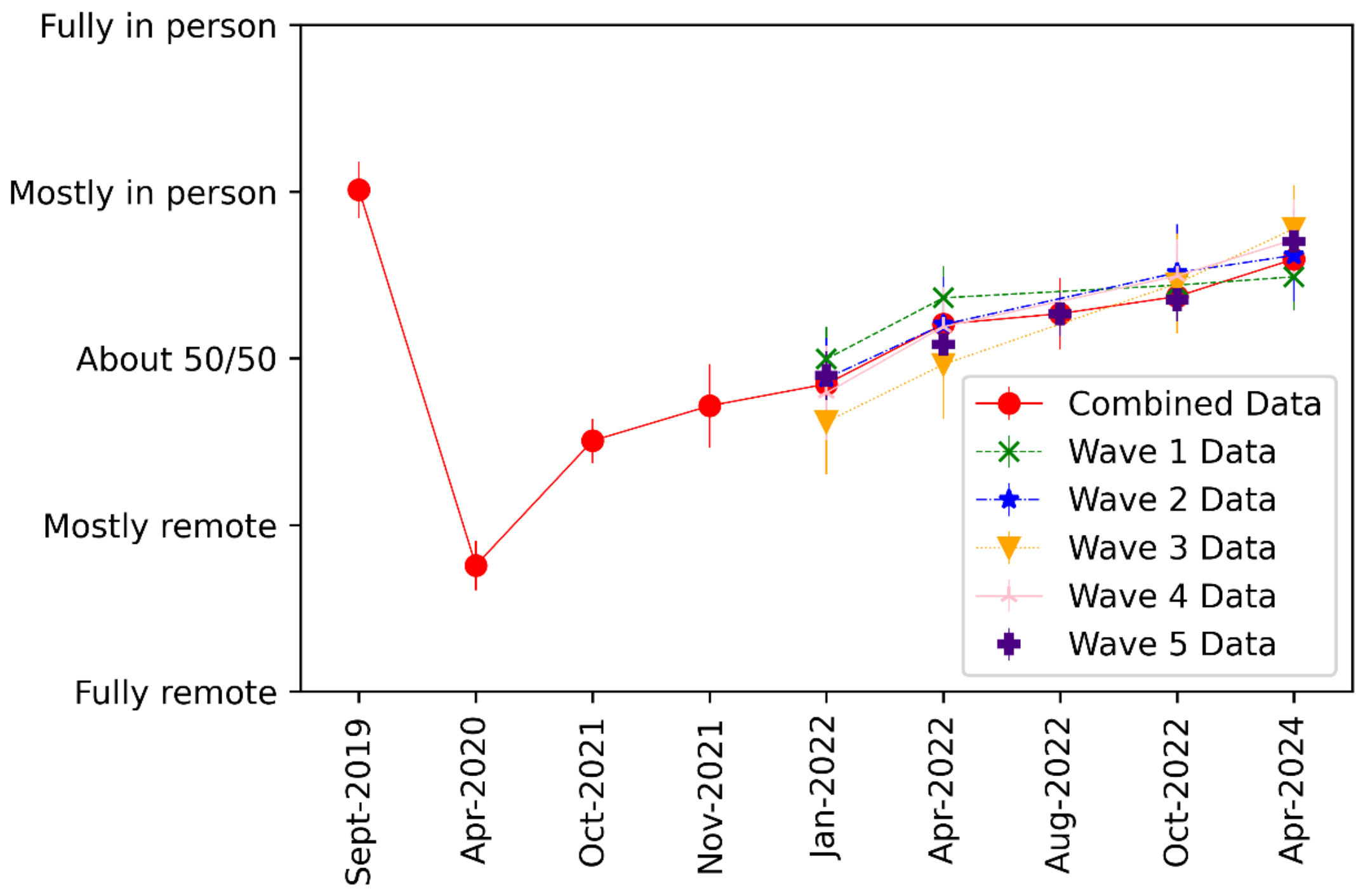

Figure 3: Average work location over time

## 5. Results

### *5.1. Employer approach to employee remote work for whom it is possible*

Figure 3 presents the average work location across organizations for employees for whom remote work is possible at different time points. The **red colored line** corresponds to a combination of data from all waves, where for respondents who participated in multiple waves, their oldest response was retained for the past points (compared to when the survey was conducted) and the most recent response was retained for the future or current time point. The lines with other colors correspond only to data collected in a particular wave and helps understand the adapting policies for a particular time point.

Several interesting insights are immediately visible from this figure. First, as expected, the average work location dropped significantly at the beginning of the pandemic from being somewhere around 'mostly in-person' in September 2019 to being somewhere between 'fully and mostly remote'. This is not surprising given the unprecedented nature of the pandemic situation, which forced employers to adopt remote work at a large scale, at least for whom it is possible. Second, as the economy opened in mid-2020 to early 2021 and mass vaccinations began, a significant rebound to in-person work happened and the average work location has increased since then. Third, despite the increasing average work location over time, in April 2024, about 4 years since the beginning of the pandemic, the average work location is between about 50/50 and mostly in-person and is still not at the pre-pandemic level, suggesting that telework will stay well beyond the pandemic is over. Fourth, our data also captures some trends in employers' policy adjustments potentially due to changing pandemic situation and employee pushback on return-to-work plans. Specifically, our data captures differences between where employers wanted their workforce to be at a future time point versus where the workforce actually was at that time point. For example, the average expected work location for wave 1 data (which was collected in October 2021) in January 2022 and April 2022 are much higher than based on data using wave 2 (collected in December 2021) and wave 3 (collected in January 2022), partly related to in-person presence roll back due to the omicron variant and also potentially related to push back from the employees regarding the return of work plans (Barrero et al., 2021a; Latifi, 2022; Noguchi, 2021).

A deeper analysis of the April 2024 data seems to suggest that most employers in our data are planning to adopt some form of remote work suggesting hybrid work practices may be the future instead of a fully remote or fully in-person approach. Specifically, only 14.2% of employers in our data reported taking a fully in-person approach, 45.8% reported a mostly in-person approach, 27.4% reported about 50/50, and 13.2% reported taking a mostly or fully remote approach. Within the last group, only one employer reported taking a fully remote approach, reinforcing that most employers will likely adopt some form of hybrid work.

Note here that while we removed the responses where for April 2024 a response of ‘wait and see’ was recorded for creating Figure 3, we conducted a sensitivity analysis to understand how the mean work location for April 2024 would vary if all those ‘wait and see’ responses decide to choose a fully in-person or fully remote response. Our analysis shows that the mean average work location would be 3.768 and 3.270 if all the ‘wait and see’ responses decide to adopt a fully in-person and fully remote approach, respectively. The corresponding mean work location for September 2019 was 4.016 which is higher (more in-person) compared to the extreme case of 3.768, strengthening the case that higher remote work is expected in April 2024.

Figure 4 presents the average work location at different time points by the employers’ sector of operations. Given the modest sample size and since over 60% of companies in our data are from transportation / manufacturing sectors, we present how the average work location evolved for the organizations which are from the transportation / manufacturing sectors compared to others (which includes sectors like information, consulting, etc.)[8]. Even though average work location was similar across the two sectors in September 2019 as well as in April 2020 (albeit with transportation / manufacturing sectors being slightly more in-person), the rebound across the two sectors has been quite different, where transportation / manufacturing / warehousing sector organizations have rebounded back to much higher in-person work compared to other organizations and these trends seems to continue in April 2024 as well. It must be noted that these trends are only for employees for whom remote work is possible, so higher in-person work in transportation / manufacturing / warehousing sectors is not due to the fact that these sectors employ more individuals that are required to be present in-person. A hypothesis is that higher in-person work in the transportation / manufacturing sector could be due to a requirement of higher coordination between those who are required to be fully in-person and those who can do their jobs remotely. Nevertheless, it is remarkable to see these differences across sectors even though all employees in question are those for whom remote work is possible.

---

[8] For figures 4 to 7, we only present results using the combined data instead of the wave specific data due to sample size constraints.

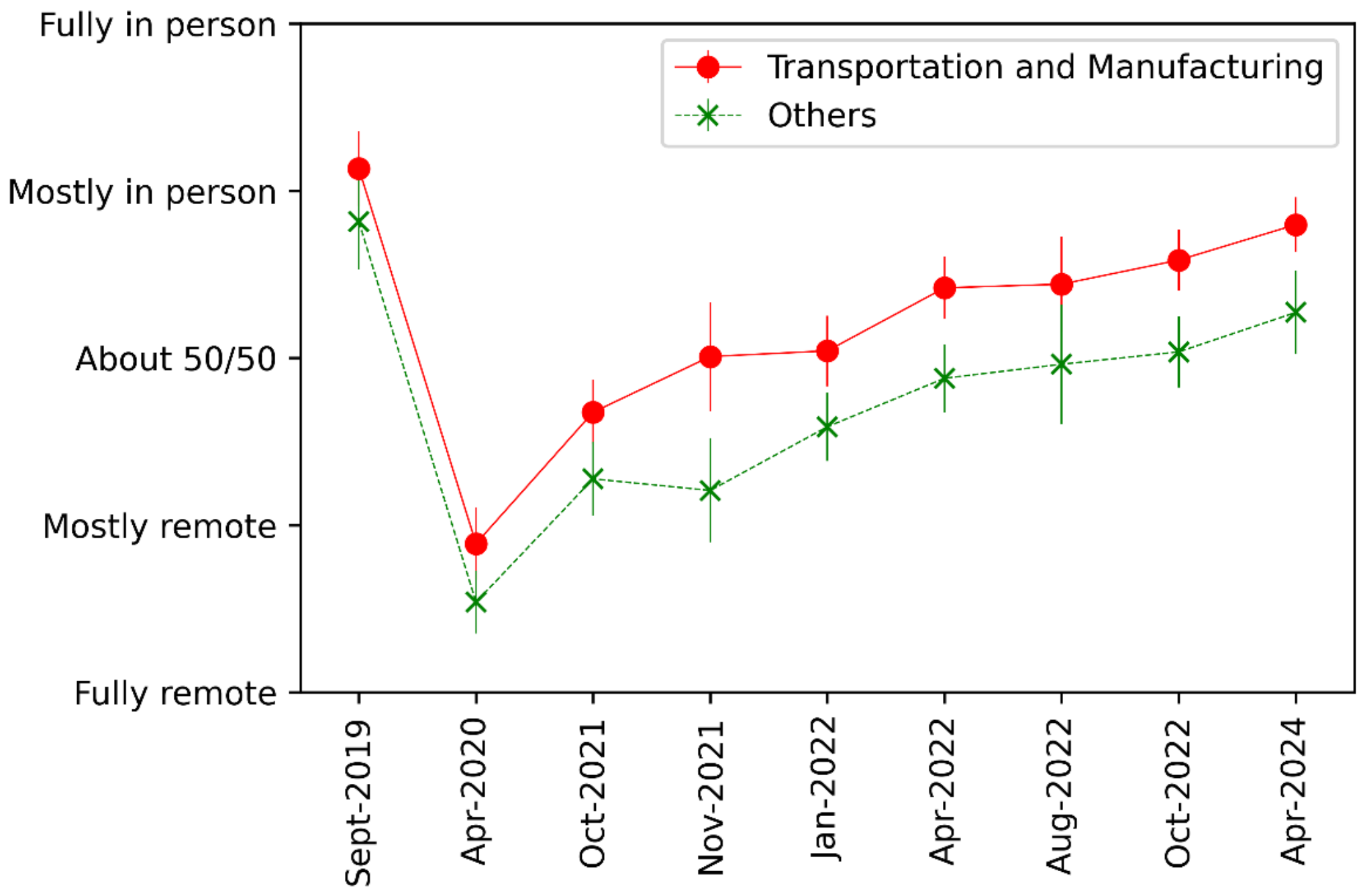

Figure 4: Average work location by sector of operations at different time points

Figure 5 presents the average work location at different time points for employees in the information technology (IT), Sales, Human Resources, Administration, Legal, and Finance departments of an organization. It is again interesting to see that even though the employers had a similar approach to remote work across the three sets of departments (with HR/Admin/Legal/Finance being slightly more in-person, followed by IT and then Sales), the gap between HR/Admin/Legal/Finance and Sales/IT has increased significantly since the pandemic with HR/Admin/Legal/Finance departments expected to be much more in-person than others. This is not surprising though since sales-related tasks can easily be done using phone or video calls and IT-related tasks also have a potential to be completed through virtual connections to IT infrastructure; and given the pandemic forced acceleration of remote work, potential cost savings for the employers and employee push back on return to work post-pandemic adoption is higher for IT and sales departments. For the HR/Admin/Legal/Finance departments, since their work potentially requires more in-person coordination, both intra-organizational and with outside parties, this leads to more in-person work. Nevertheless, it must be noted that the average work location in April 2024 is still below the pre-pandemic trends in September 2019 for all three department groups suggesting a companywide increase in remote work adoption.

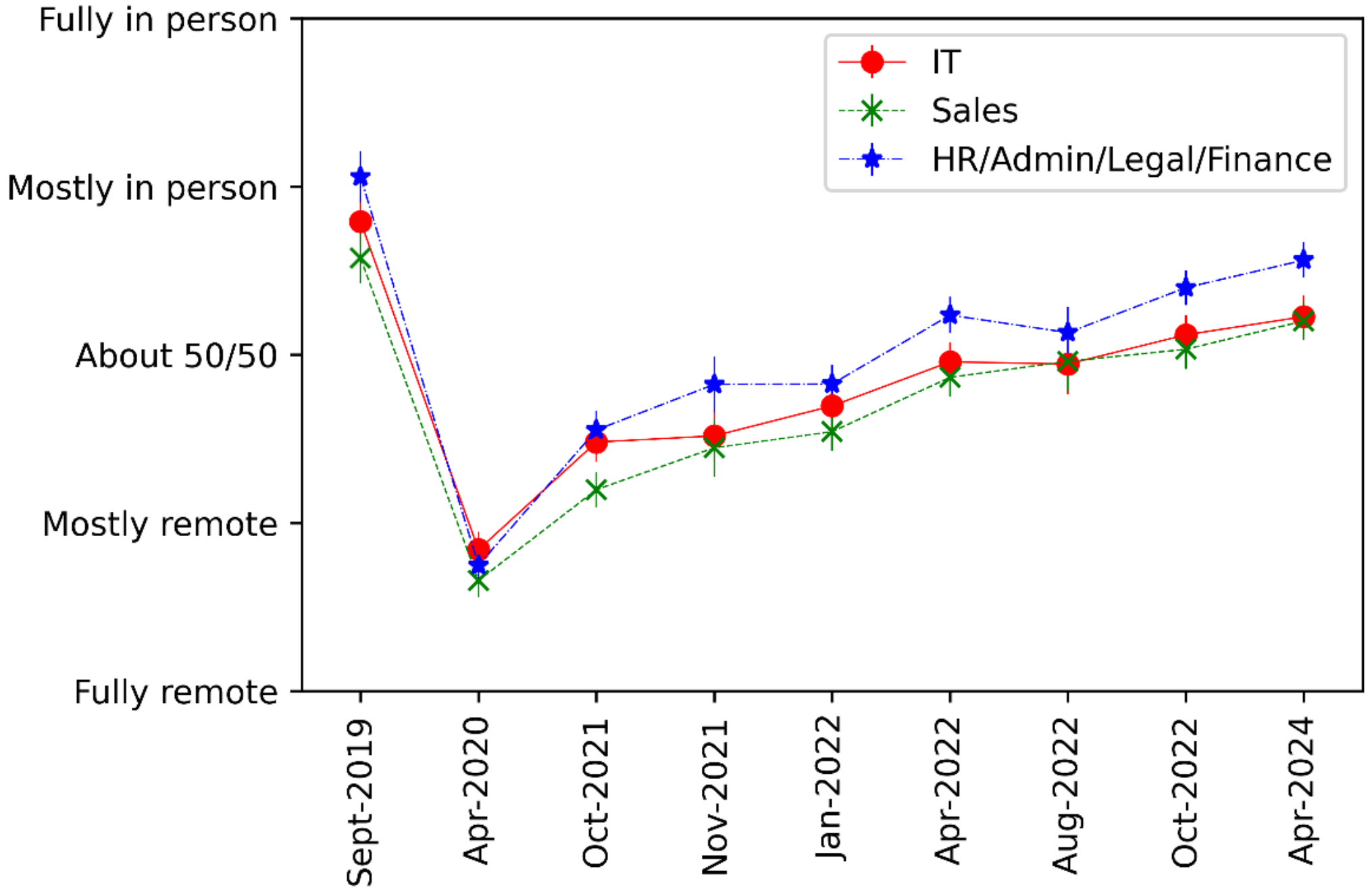

Figure 5: Average work location by various departments

Figure 6 presents the average work location of various organizations by their pre-COVID work location approach. This data provides several insights into how the remote work landscape has changed for organizations with different remote work approach pre-COVID. First, all organizations irrespective of their pre-COVID approach to remote work heavily adopted remote work during the early period of the pandemic in April 2020. This is understandable since the unprecedented nature of the pandemic forced most employers to adopt remote work. Second, the results also show that the rebound back to higher in-person work was and is expected to be higher for those who were fully in-person pre-COVID compared to others. Third, it is interesting to note that the average work location for those who were in the *others* category pre-COVID has rebounded higher than their pre-COVID average work location indicating that not all organizations have seen an increase in remote work as a result of the pandemic, some have seen a reverse impact as well. A deeper investigation of this data shows that there were 24 organizations in this category with 11 in transportation / manufacturing and 13 in other categories, however, almost all of them were closely related to the transportation sector like railroads, freight, logistics, or transportation-related data-based organizations. Given that these sectors have considerably grown during the pandemic (partly attributable to e-commerce) and also faced higher performance pressure due to global supply chain disruption, an increase in in-person work during and beyond the pandemic for these organizations is potentially due to these factors.

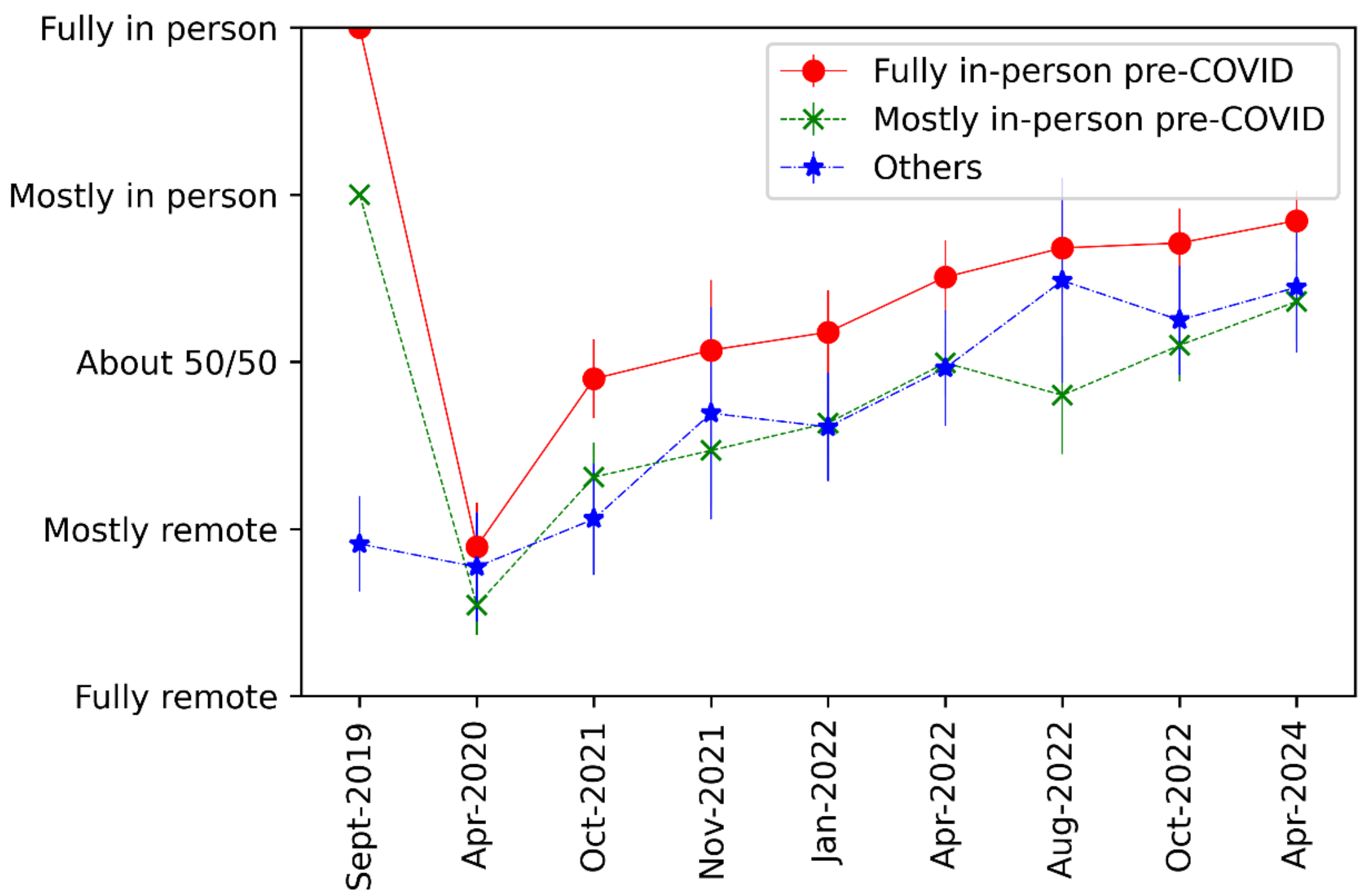

Figure 6: Average work location by pre-COVID remote work policies

Lastly, Figure 7 presents the average work location of various organizations by their April 2020 work location approach. This data helps us understand whether a pandemic-forced shift in work location policies impacted the organizations' long-term policies. In this regard, a key finding is that those with higher remote work adoption in April 2020 maintained higher remote work throughout the pandemic than others and are also likely to maintain these trends in April 2024 (though the difference is perhaps not significant). An interesting observation here is that those who were fully remote in April 2020 were slightly more in-person pre-COVID than those with a mostly remote approach during April 2020. This potentially indicates a sort of cultural shift in some organizations who shifted to fully remote work at the height of the pandemic and maintained a higher remote work through the pandemic compared to others.

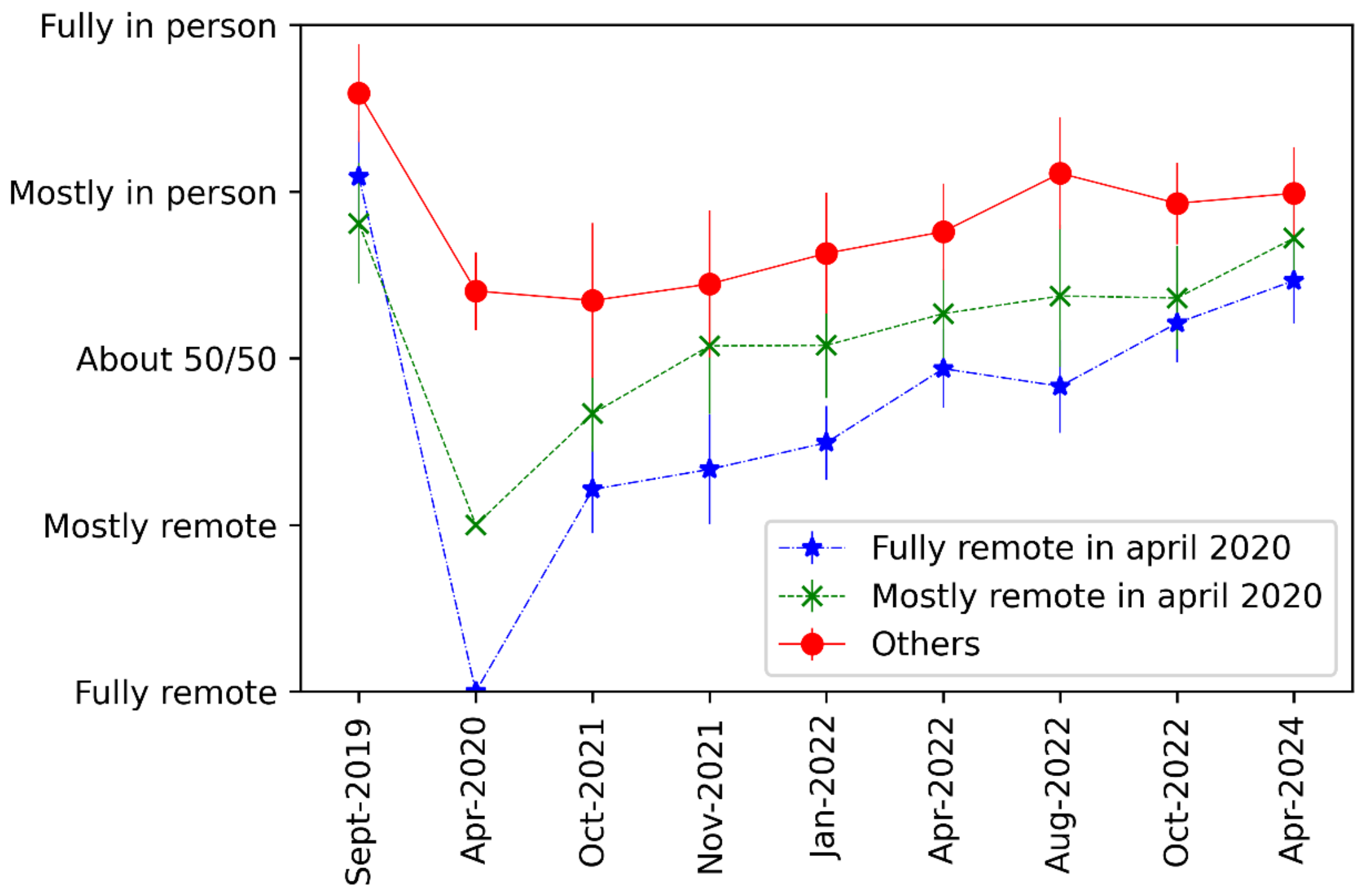

Figure 7: Average work location by April 2020 work location policies

Overall, several key insights emerged from this analysis. First, our data suggest that the pandemic accelerated remote work adoption is likely to stick well beyond the pandemic with some form of hybrid work likely to be the norm for most organizations. Second, there is a significantly higher remote work adoption pattern between transportation / manufacturing sectors compared to others; and it is also higher for sales and IT departments, compared to those in HR / Legal / Administration / Finance departments. This is interesting since all this data correspond to those for whom remote work is possible, indicating a potential requirement for coordination between employees for whom remote work is possible and those who are required to work in-person. Third, the post-pandemic in-person work extent is expected to be higher for those working fully in-person work pre-COVID. However, not all organizations have seen an increase in remote work as a result of the pandemic, some have seen a reverse impact as well, potentially related to business growth during the pandemic.

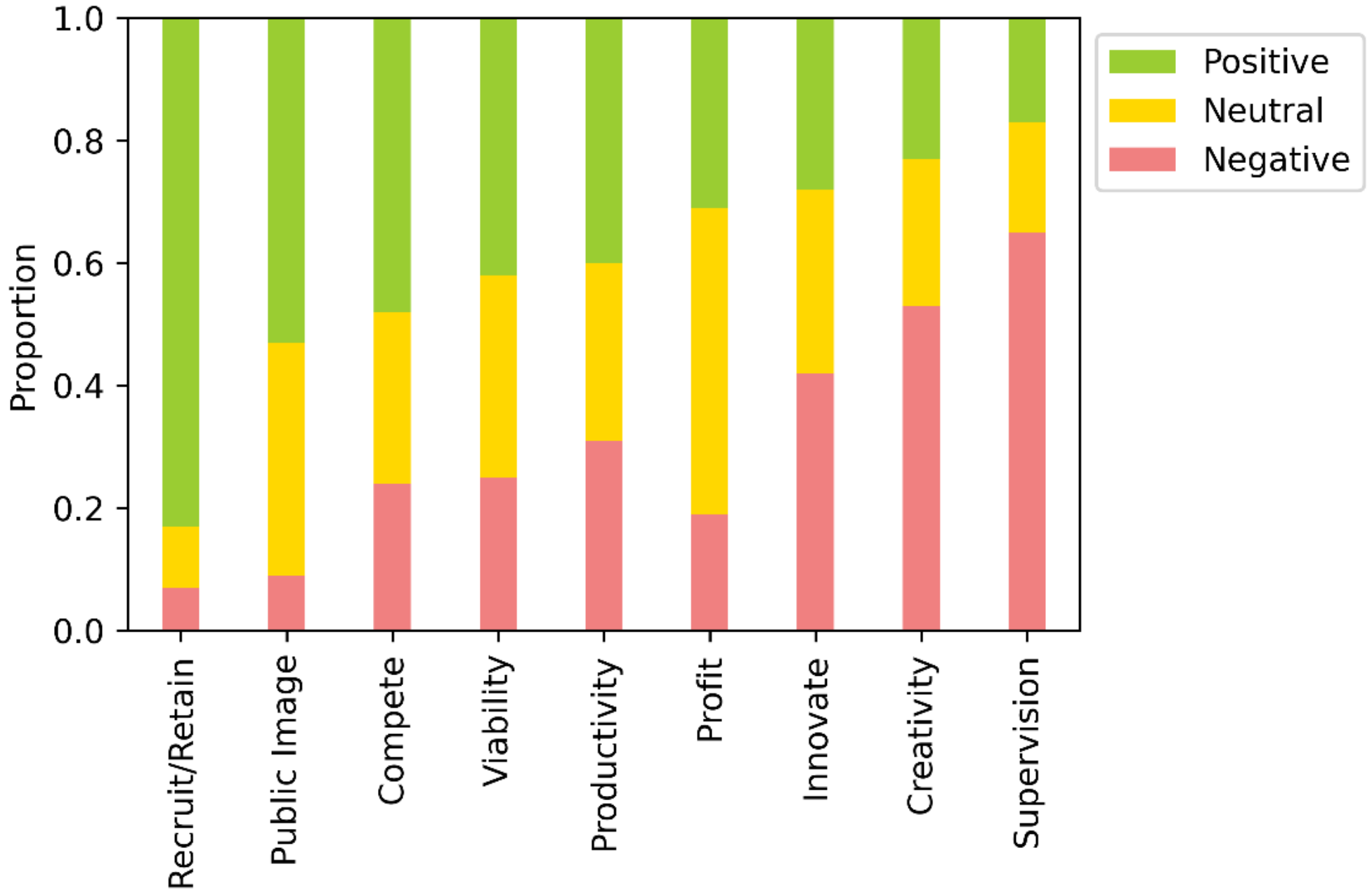

Figure 8: Employer opinion of the impact of 2-days a week remote work policy on various business aspects

*5.2. Employer opinion regarding the impact of remote work on various business aspects*

*Top benefits and concerns related to remote work.* Figure 8 presents descriptive statistics from the 9 response items related to employer opinion regarding the impact of 2-days a week remote work program on various business aspects. Several interesting insights can be derived from this figure. First, the ability to supervise/mentor, the effect on creativity and innovation are the aspects where employers see the most negative impact of remote work. On the other hand, a majority of employers agree that a 2-days a week remote work policy will have a positive impact on their ability to recruit and retain employees, their public image, and their ability to compete. Lastly, business viability and profit and employee productivity are in the middle. Regarding the aspects with the most negative impact, our results align with existing research that remote work may lead to an adverse impact on innovation and creativity (Yang et al., 2022), so, naturally, it is an important factor that employers are considering. Our results on the positive impact of remote work on the ability to recruit / retain also aligned with the emerging literature regarding employees demanding more flexibility (Barrero et al., 2021a; Dua et al., 2022; Owl Labs, 2022; Parker et al., 2020). It is interesting to see most employers reporting a neutral impact on profit, while the employers seem to be divided on the aspect of employee productivity.

*Latent Class Analysis*. To gain a deeper understanding of these results, we use this data to estimate a latent class model with and without covariates. Figure 9 presents the item response probabilities ($\rho$) for the two latent classes we found and for 9 items and three response categories. Table 2 presents the results from the membership model and the estimated population shares ($\gamma$) for each latent class. Several insights emerge from this analysis. First, we name the two latent classes as employers with a positive (class 1) or negative (class 2) outlook towards remote work based on the item response probability values for each item. Specifically, for class 1, the probability of a positive response is higher than for class 2, for most response items, hence it makes sense for these employers to be termed as those with a positive outlook towards remote work. Similarly, for class 2, the probability of a negative response is higher than for class 1 for most responses, hence it makes sense for these employers to be terms as those with a negative outlook towards remote work.

Table 2: Latent Class Membership Model and Population Shares

| Variable | Parameter Estimate | t-stat |
|---|---|---|
| Constant | -0.262 | -1.045 |
| Transportation and Manufacturing sector indicator | 0.958 | 2.152 |
| Model Fit | | |
| Number of observations | 129 | |
| Log-Likelihood | -867.952 | |
| Estimated population shares ($\gamma$) | | |
| Class 1: Positive Outlook | 0.525 | |
| Class 2: Negative Outlook | 0.475 | |

In our data, 52.5% of employers belong to the positive outlook class and 47.5% belong to the negative outlook class. Based on the item response probabilities, variables where the employers are most divided include the ability to supervise / mentor, impact on innovation and creativity (where employers in class 2 see remote work to have a highly negative impact while those in class 1 are almost equally likely to report positive, neutral or negative impact). Further, employers in both classes seem to agree that remote work policies have a positive impact on the ability to recruit / retain. The results from the membership model estimation suggest that those in Transportation / Manufacturing sectors were more likely to have a negative outlook toward the impact of remote work on various business aspects compared to other sectors potentially suggesting a large interaction between the nature of work, level of coordination required between in-person employees and those who can work remotely and the employers' perception of the impact of remote work.

### *5.3. Future Landscape of Remote Work*

Table 3 presents the results from the ordered probit model of the April 2024 work location approach, with the dependent variable on a 4-point ordered scale (1 = fully or mostly remote, 2 = about 50/50, 3 = mostly in-person, 4 = fully in-person). Here a positive parameter corresponds to a higher likelihood of in-person work in April 2024 compared to the base group. We determined the final specification for this model after extensive testing and retained only four variables in the final specification. Admittedly, two less significant variables were also retained due to their intuitive interpretation and the smaller size of our sample.

Four key insights emerged from this analysis. First, those in the transportation / manufacturing / warehousing sectors are significantly more likely to maintain a higher in-person presence than those in the other sectors. This aligns with the results presented earlier and potentially relates to the nature of the work, which required higher coordination between those who work fully in-person and those who have the option to work remotely. Second, our results also show that those who were fully in-person pre-COVID are more likely to be in-person in April 2024. Interestingly, this variable was significant even after controlling for the sector of operations, indicating that past remote work approach in an indicator of future approach in the other sectors, again potentially related to the nature of work done by an organization. The organizations in this segment (other sectors with a fully in-person approach pre-COVID) in our data included a few public agencies and some engineering consulting firms. Given that such organizations typically have tasks that requires interactions with other involved parties like clients, public officials, construction sites crews, etc., it is intuitive that these organizations are expected to take a higher in-person work approach in April 2024. Third, those who were fully remote in April 2020 are marginally less likely to be in person in April 2024. This could potentially be related to a cultural shift in the remote work approach within on organization where an employer decided to continue a higher remote work approach after a positive experience in the early phase of the pandemic; or could also be a result of employee pushback to return to work. Lastly, our results show that those with a negative outlook toward remote work are more likely to be in-person in the future. Note here that this variable is insignificant in this model potentially due to a strong association with the sector of operation variable as revealed by the membership model presented earlier. However, when we re-estimated this model without the sector of operations variable, the parameter corresponding to the probability of being in the negative outlook class was significant.

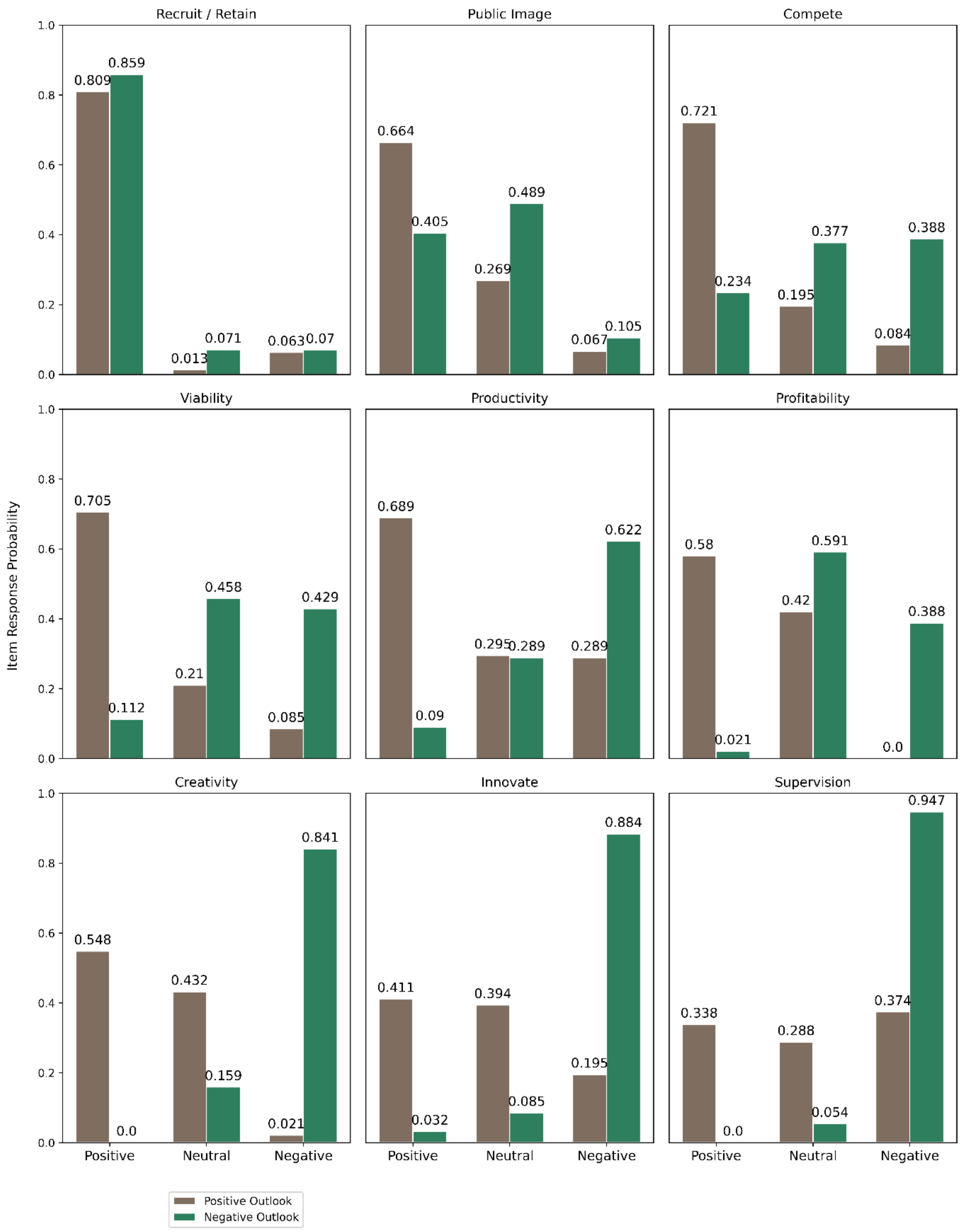

Figure 9: Item response probabilities for various business aspects for the two identified clusters

Overall, a key takeaway here includes a high interaction between the nature of work, sector of operations and potential impact remote work may have on business as leading factors impacting future remote work policies.

Table 3: Ordered probit model of April 2024 work location

| Variable | Parameter Estimate | T-Statistic |
|---|---|---|
| Transportation, Warehousing, and Manufacturing sector indicator | 0.492 | 1.916 |
| Fully in-person approach pre-covid indicator | 0.525 | 2.352 |
| Fully remote approach in April 2020 indicator | -0.281 | -1.248 |
| Probability of being in the negative outlook class | 0.302 | 1.126 |
| **Thresholds** | | |
| 1\|2 | -0.637 | -2.637 |
| 2\|3 | 0.343 | 1.558 |
| 3\|4 | 1.769 | 7.852 |
| Fit Measure | 0.170 | |
| No. of observations | 105 | |

### 5.4. *Business travel, in-person client interactions, office space reorganization and work arrangements employers are willing to consider*

#### 5.4.1. Business travel and in-person client resumption compared to pre-pandemic

Figure 10 presents the percentage of business travel of over 50 miles and in-person client interaction that has returned compared to the pre-pandemic levels at the time of different waves of data collection. Regarding business travel, our data suggests that the business travel trends were at 32% in October 2021, 40% in December 2021, 36% in January 2022, 48% in April 2022, and 55% in August 2022 compared to pre-pandemic levels. It is interesting to note that the trends in August 2022 were still well below the pre-pandemic levels and potentially indicating a slower rebound. Our numbers on business travel are also close to the estimates from a survey by the Global Business Travel Association which found out that the domestic business was at about 63% of pre-pandemic levels in early October 2022 (GBTA, 2022). It would be interesting to see how business travel recovers from here, especially given that there are growing concerns regarding a slowing economy amongst employers (Levere, 2022). Regarding in-person client interaction, the trends were at 30% in October 2021, 43% in December 2021, 32% in January 2022, 51% in April 2022 and 61% in August 2022 compared to pre-pandemic trends. It is not surprising that the recovery of local in-person client interactions is slightly faster than business travel, potentially due to the involved cost as well as higher (perceived) contagion risk when flying. Another interesting trend in our data is the reduced business travel and in-person client interactions around January 2022 compared to the previous month, potentially due to the changing pandemic landscape due to the omicron variant.

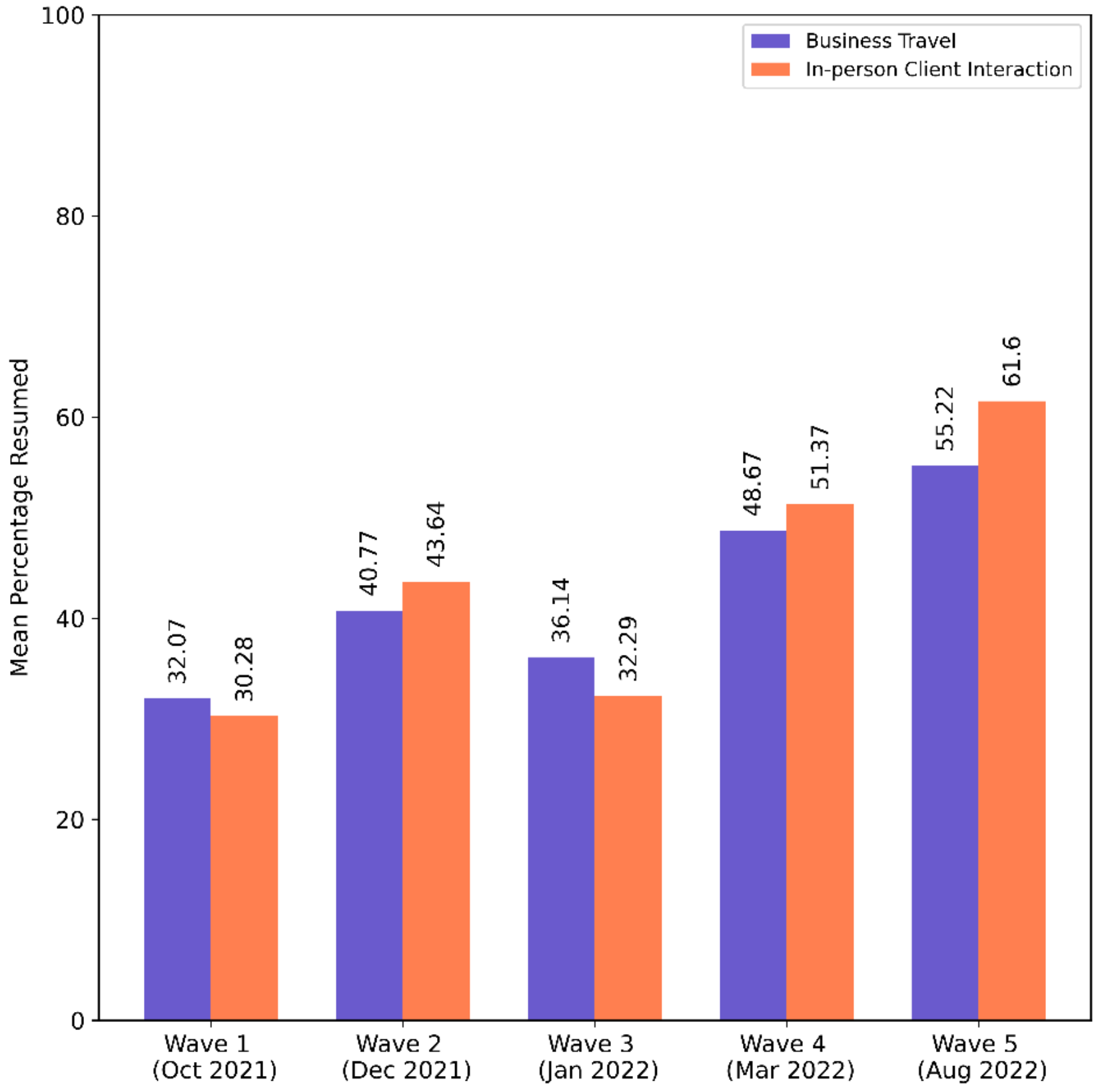

Figure 10: Percent of business travel and in-person client interactions returned during various waves of data collection

### 5.4.2. Office space reorganization

To gain a better understanding of the permanency of changes in the work location landscape, we also asked the employers whether they have or plan to relocate, expanded, or reduced office spaces since the beginning of the pandemic. Figure 11 shows the number of employers with various office space reorganizations adopted (or to be adopted) by them. This question was only asked in Wave 5 (August 2022) where 38 out of 56 respondents reported making some changes to their office space (i.e. ~32% made no changes). The most reported response was a reconfiguration of office space to cater to changing nature of the work environment (21/38), followed by reduced office space (10/38). For others, there seems to be mixed response of either increase or decrease in office space in same the same or different building or area. We also asked the respondents to self-describe the nature of their office same reorganization in a few words. Based on these open-ended responses, a general trend was that those who expanded or are planning to expand soon are growing companies, however, their growth in office space has been slower than expected due to changing work location landscape.

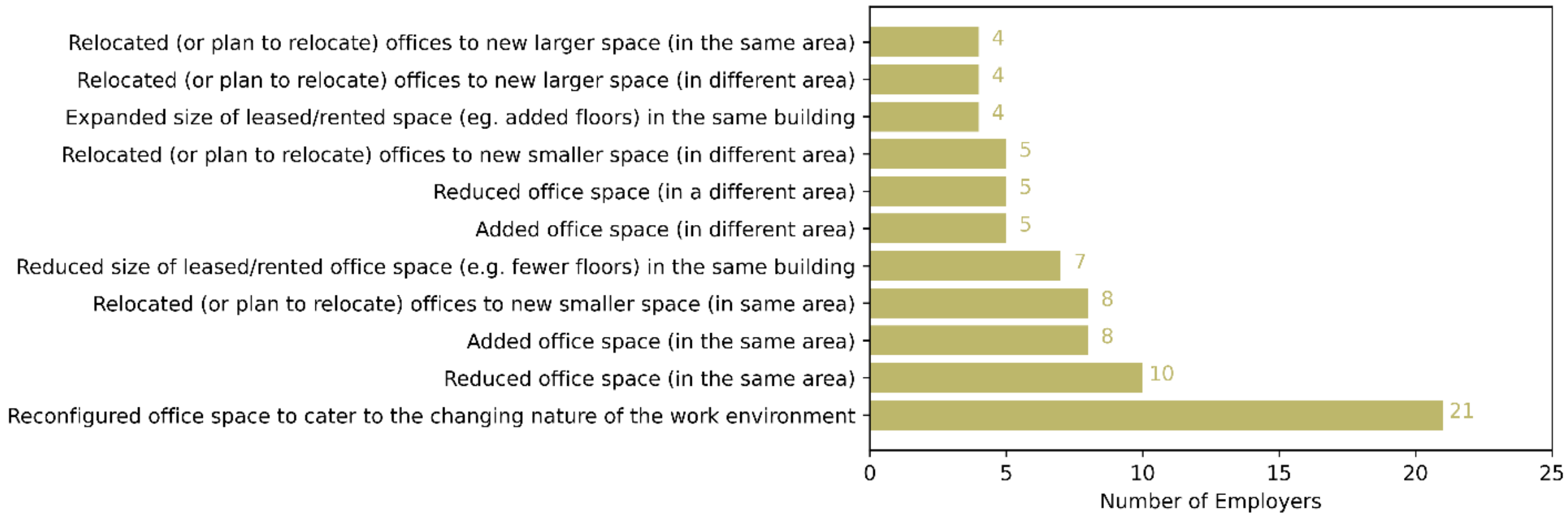

Figure 11: Office space reorganization made by the employers since the beginning of the pandemic

## 6. Summary, Key Takeaways, Limitations, and Future Work

### *6.1. Summary and Key Findings*

Using data from top executives from 129 employers in North America from a 5-wave longitudinal survey, this study presents an employer-side perspective on telework through and beyond the pandemic. Specifically, we present data on how employer approach to remote work evolved over time and what is expected in April 2024, four years since the beginning of the pandemic. We also identify employers' top concerns and expected benefits related to remote work, which will potentially shape their future remote work policies. To dig a little deeper, we conduct a latent class analysis which divides the employers into latent classes based on their outlook towards remote work. Next, we present results from an ordered probit model to understand the factors that may shape employers' April 2024 remote work approach and then present results on resumption of business travel of over 50 miles and in-person client interactions at the time of the 5 waves of data collection, followed by how employers' have or expect to reconfigure their office spaces in the future.

The following key insights emerged from the above analyses. First, our data suggests that the pandemic accelerated remote work adoption is likely to stick well beyond the pandemic with some form of hybrid work likely to be the norm for most organizations. We also found out that there is a much higher rebound to in-person work in the transportation / manufacturing / warehousing sectors, compared to other sectors; and this trend also exists for employees in HR / Legal / Administration / Finance departments compared to Sales / IT departments. This is interesting since all this data correspond to those for whom remote work is possible, indicating a potential requirement for coordination between employees for whom remote work is possible and those who are required to work in-person. Our results also show that the post-pandemic in-person work extent is expected to be higher for those working fully in-person work pre-COVID. However, not all organizations have seen an increase in remote work as a result of the pandemic, some have seen a reverse impact as well, potentially related to business growth during the pandemic.

Second, on the end of employers' opinion regarding impact of remote work on various business aspects, the ability to supervise and mentor, and the effect on creativity and innovation are the aspects where employers see the most negative impact of remote work. On the other hand, a majority of employers agree that a 2-days a week remote work policy will have a positive impact on their ability to recruit and retain employees, their public image, and their ability to compete. Lastly, business viability and profit and employee productivity are in the middle. Most employers report a neutral impact on profit, while employers seem to be divided on the aspect of employee productivity. Based on a Latent Class analysis, we divide the employers into two latent classes: those with a positive outlook on the impact of remote work on business and those with a negative impact of remote work on business. variables where the employers are most divided include the ability to supervise / mentor, the impact on innovation and creativity (where employers in class 2 see remote work to have highly negative impact while those in class 1 are almost equally likely to report positive, neutral or negative impact). Further, employers in the both classes seem to agree that remote work policies have a positive impact on ability to recruit / retain. The results from the membership model estimation suggest that those in Transportation / Manufacturing sectors were more likely to have a negative outlook towards the impact of remote work on various business aspects compared to other sectors potentially suggesting a large interaction between the nature of work, level of coordination required between in-person employees and those who can work remotely and the employers' perception of the impact of remote work.

Third, based on the results from the estimated ordered probit model of April 2024 work location approach, our results indicate that those in transportation /warehousing / manufacturing sectors, those with a fully in-person approach to remote work, and those with a negative outlook towards the impact of remote work on business are likely to be more in-person going forward and those with fully remote work in April 2020 are less likely to be fully in-person.

Lastly, we also found out that as of August 2022, about 55% business travel of over 50 miles and 61% of in-person client interaction has resumed compared to pre-COVID level, indicating a slower rebound to pre-pandemic level of work-related in-person travel. On the end of office space reconfiguration, we find mixed results with over 30% of employers reporting no changes to their office spaces. Amongst those who made some form of changes, a reconfiguration of the office space to cater to changing nature of work was the most popular approach with some employers also reporting an expansion due to a business growth. However, those who grew reported the office space expansion to be less than expected due to changing work location landscape.

A few limitations are worth mentioning here. First, our sample size is admittedly small and a future study with a larger sample size would be of great value to gain a thorough understanding of changing remote work landscape. Second, our data over-represents the transportation / warehousing / manufacturing sector, which prevented us from gaining a deeper understanding of the difference between work location practices in the other sectors. Third, the work location landscape continues to evolve and the true future of work is potentially going be a function of several factors and thus required continued monitoring of the situation.

Overall, it is evident that the work landscape continues to evolve for most companies, as these struggles to adapt to changing employee preferences and expectations at the same time as customer demand patterns for goods and services continue to shift, and global supply chains endeavor to overcome setbacks related to various external events (wars, energy prices, critical component shortages, inflationary environment, etc.). From a transportation planning standpoint, it is essential to consider various possible scenarios of hybrid work arrangements and associated impacts on office occupancy and joint work and residential shifts, and seek the sweet spot between healthy and rich urban environments on one hand, and the potential reduction in peak commuting travel made possible by remote work. Continuing surveys along the lines presented in this paper can inform the shape and likelihood of such scenarios and help guide related policies.

## Acknowledgments

Partial funding for the research on which this paper is based is provided the U.S Department of Transportation Tier I University Transportation Center on Telemobility awarded to Northwestern University in partnership with University of California, Berkeley and the University of Texas, Austin. Partial support received by the first author through Northwestern University Transportation Center's Dissertation Year Fellowship is gratefully acknowledged. The contents remain the sole responsibility of the authors and do not necessarily reflect the positions of the sponsoring agency.

## CRediT authorship contribution statement

The authors confirm contribution to this paper as follows: study conception and design: DT, HM, AS, MS, SS, JW, BJ; data collection: DT, HM, AS, MS, SS, JW, BJ; analysis and interpretation: DT, HM, AS, MS, SS, JW; draft manuscript preparation: DT, HM, AS, MS. All authors reviewed the results and approved the final version of the manuscript.